\documentclass[a4paper,10pt,5p,times,twocolumn,superscriptaddress,floatfix,review,preprint]{elsarticle}
\biboptions{numbers,sort&compress}
\usepackage{lineno,hyperref}
\usepackage{epsfig,graphicx}
\usepackage{amstext}
\usepackage{amsmath}
\usepackage{amssymb}
\modulolinenumbers[5]
\allowdisplaybreaks[4]

\journal{Fundamental Research}

\begin{document}
	
\title{Optomechanical Schr\"{o}dinger cat states in a cavity Bose-Einstein condensate}

\author[address1]{Baijun Li\fnref{myfootnote}}

\author[address2]{Wei Qin\fnref{myfootnote}}

\author[address1,address3]{Ya-Feng Jiao\fnref{myfootnote}}

\author[address1] {Cui-Lu Zhai}

\author[address1]{Xun-Wei Xu}

%\author[address1]{Jie-Qiao Liao}

\author[address1]{Le-Man Kuang\corref{correspondingauthor}}
\cortext[correspondingauthor]{Corresponding authors.}
\ead{lmkuang@hunnu.edu.cn}

\author[address1]{Hui Jing\corref{correspondingauthor}}
\ead{jinghui73@gmail.com}

\address[address1]{Key Laboratory of Low-Dimensional Quantum Structures and Quantum Control of Ministry of Education, \\
Department of Physics and Synergetic Innovation Center for Quantum Effects and Applications, \\
Hunan Normal University, Changsha 410081, China}

\address[address2]{Theoretical Quantum Physics Laboratory, RIKEN Cluster for Pioneering Research, Wako-shi, Saitama 351-0198, Japan}

\address[address3]{Laboratory of Chemical Biology $\&$ Traditional Chinese Medicine Research, \\
Ministry of Education College of Chemistry and Chemical Engineering, \\
Hunan Normal University, Changsha 410081, China}

\fntext[myfootnote]{These authors contributed equally to this work.}

\begin{frontmatter}
	
\begin{abstract}
Schr\"{o}dinger cat states, consisting of superpositions of macroscopically distinct states, provide key resources for a large number of emerging quantum technologies in quantum information processing. Here we propose how to generate and manipulate mechanical and optical Schr\"{o}dinger cat states with distinguishable superposition components by exploiting the unique properties of cavity optomechanical systems based on Bose-Einstein condensate. Specifically, we show that in comparison with its solid-state counterparts, almost a $3$ order of magnitude enhancement in the size of the mechanical Schr\"{o}dinger cat state could be achieved, characterizing a much smaller overlap between its two superposed coherent-state components. By exploiting this generated cat state, we further show how to engineer the quadrature squeezing of the mechanical mode. Besides, we also provide an efficient method to create multicomponent optical Schr\"{o}dinger cat states in our proposed scheme. Our work opens up a new way to achieve nonclassical states of massive objects, facilitating the development of fault-tolerant quantum processors and sensors.
\end{abstract}	

\begin{keyword}
Schr\"{o}dinger cat state, Cavity Optomechanics, Bose-Einstein condensate, Collective density excitation, Squeezing.
\end{keyword}

\end{frontmatter}
	
%\linenumbers
	
\section{Introduction} \label{Int}
Schr\"{o}dinger cat states, characterized by the quantum superpositions of macroscopically distinct states, are the peculiar features of the quantum systems. This type of nonclassical state has solid theoretical and experimental foundations, and hence it has become a commonplace over the past years. Cat states are not only of crucial importance for fundamental researches in quantum physics, e.g., conceptually exploring the boundaries between the classical and quantum domains~\cite{Zurek03Decoherence}, but also provide pivotal fault-tolerant quantum resources for applications ranging from quantum information processing~\cite{Ku20Experimental} to quantum sensing~\cite{Pezze18Quantum}. So far, cat states have been demonstrated and manipulated in a myriad of mesoscopic-scale physical platforms, including those involving photons~\cite{Ourjoumtsev07Generation,Ourjoumtsev06Generating,Nielsen06Generation,Vlastakis13Deterministically,Kirchmair13Observation,Sychev17Enlargement}, superconducting cavities~\cite{Leghtas15Confining,Grimm20Stabilization}, ions, and atoms ~\cite{Monroe96A,Myatt00Decoherence,Leibfried05Creation,Omran19Generation,Lu19Global,Song19Generation}. Very recently, by utilizing the radiation pressure induced coherent light-motion interactions, plenty of proposals have also been proposed for generating cat states of massive mechanical oscillators in cavity-QED systems~\cite{Chen21Shortcuts,Sun21Remote,Qin21Generating,Bose97Preparation,Tan13Generation,Liao16Macroscopic,Qin19Proposal,Khazali18Progress}, such as cavity optomechanical (COM) systems~\cite{Bose97Preparation,Tan13Generation,Liao16Macroscopic,Qin19Proposal} and cavity magnomechanical devices~\cite{Sun21Remote}. However, in terms of the current experimental techniques of solid-state COM systems, it still remains a big challenge to reach the ultrastrong single-photon COM coupling regime~\cite{Aspelmeyer14}, i.e., $g/\omega_b>1$, [cf.\,Eq.\,(\ref{Qumham})], whereby it further hinders the creation of cat states with distinguishable superposition components~\cite{Liao16Macroscopic}.

Here we propose how to generate and manipulate mechanical and optical cat states in a Bose-Einstein condensate (BEC)-based COM system that consists of a high-finesse Fabry-Perot cavity and an optically trapped one-dimensional BEC, unveiling its unique properties that are otherwise unattainable within the solid-state COM devices. We note that the BEC-based COM systems have already been extensively studied in experiments~\cite{Murch08Observation,Brennecke08Cavity,Smith11Optomechanical,Spethmann16Cavity,Purdy10Tunable}, providing a promising platform for not only achieving nonclassical states of macroscopic objects, but also investigating many kinds of nonlinear COM effects such as mechanical ground-state cooling~\cite{Smith11Optomechanical}, quantum simulating~\cite{Jing11Quantum,Kumar21Cavity}, quantum-state transfer~\cite{Singh12Quantum}, and multistable oscillating~\cite{Dong11Multistability}. A recent advance closely related to our study is, for example, the exploration of a role reversal BEC-based COM system by Zhang \textit{et al.}~\cite{Zhang12Role}, in which they analyzed several nonclassical states of the atomic matter-wave field, and particularly an atomic cat state. As revealed in these theoretical and experimental researches, the advantages of such BEC-based COM systems, compared with its solid-state counterparts, are twofold~\cite{Murch08Observation,Brennecke08Cavity,Smith11Optomechanical,Spethmann16Cavity,Purdy10Tunable}: (i) the BEC, whose collective density excitation serves as the mechanical mode, could be prepared in its ground state, thus leading to the situations where there is no thermal phonon occupancy presented; (ii) the strong or even ultrastrong coupling regime between photons and phonons could be achieved, whereby a single-photon excitation induced mechanical displacement could be clearly distinguished from the zero-point fluctuation~\cite{Liao16Macroscopic}.

In this paper, we further show that these unique properties of the BEC-based COM systems have great benefits in the generation and manipulation of mechanical and optical cat states. Specifically, we find that in comparison with solid-state COM systems, almost a $3$ order of magnitude enhancement in the size of a Schr\"{o}dinger mechanical cat state could be achieved for BEC-based COM systems. Correspondingly, the resulting overlap between the two superposed coherent-state components of this cat state simultaneously becomes smaller, which is quite useful for quantum information processing. Furthermore, we also show that by making use of this generated mechanical cat state, an efficient quadrature squeezing can be achieved for this mechanical mode. Moreover, in contrast to the familiar situations in the solid-state COM systems, the orientations of this quadrature squeezing could be straightforwardly engineered via tuning the Feshbach resonance of the magnetic control field~\cite{Inouye98Observation}, thereby providing a new degree of freedom to manipulate such mechanical quadrature squeezing. In addition, by altering the choice of system parameters and initial conditions, we also demonstrate that our proposed scheme can enable the generation and manipulation of multicomponent optical cat states as well. These results, as an example to achieve specific macroscopic quantum state, provide an efficient way to investigate the nonclassical nature of massive objects, facilitating the exploration of various ultrastrong-coupling-enabled quantum effects, such as photon or phonon blockade~\cite{Shahmoon20Cavity}, and quantum phase transition~\cite{Baumann10Dicke}. Moreover, apart from the one-dimensional BEC-based COM system, our work can also be realized with other types of BEC gases, e.g., annular BEC~\cite{Kumar21Cavity}, spinor BEC~\cite{Jing11Quantum,Stenger98Spin}, multi-component BEC~\cite{Myatt97Production,Kuang07Generation} and molecular Fermi-BEC~\cite{Greiner03Emergence}. As such, we believe that our proposed scheme is poised to serve in fundamental tests of quantum theories, provide key quantum resources to emerging quantum technologies, and function as promising platforms to investigate the coherent light-motion interactions.

\begin{figure*}[t]
	\centerline{\includegraphics[width=0.99\textwidth]{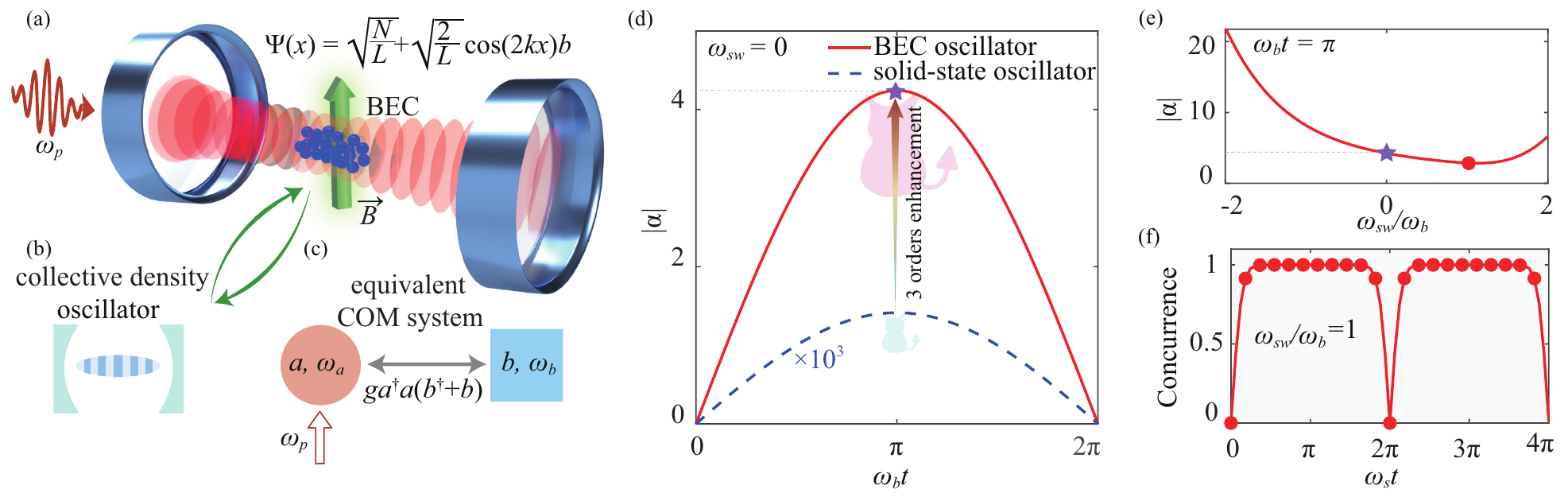}}
	\caption{(a-c) A BEC of $N=1.2\times10^{5}$ Rb atoms in the ground state is trapped inside an optical ultrahigh-finesse Fabry-Perot cavity. The collective density excitation of the condensate can act as the BEC mechanical oscillator, which is coupled to the cavity field strongly. (d) Coherent amplitude of mechanical cat states $|\alpha|$ as a function of the scaled time $\omega_{b} t$ in the solid-state and BEC-based COM systems, respectively. Compared to the solid-state COM system (blue dashed curve, $|\alpha|\approx10^{-3}$~\cite{Verhagen12Quantum}), our cat state size can be increased by three order of magnitude. (e) Coherent amplitude $|\alpha|$ in the BEC-based COM system versus the $s$-wave scattering frequency $\omega_{sw}$ at a fixed scaled time $\omega_{b} t = \pi$. (f) Time dependence of the concurrence of the state $|\psi(t)\rangle$ with the $s$-wave scattering frequency $\omega_{sw}/\omega_{b}=1$. The simulation parameters used for the solid-state COM system are: the COM coupling strength $g=10^4\mathrm{Hz}$, and the mechanical frequency $\omega_b=10^7\mathrm{Hz}$~\cite{Verhagen12Quantum}.  The used parameters for the BEC-based COM system are $\omega_{b}= 2\times10^5~\mathrm{Hz}$~\cite{Spethmann16Cavity} and $g=6\times10^5~\mathrm{Hz}$~\cite{Murch08Observation}.} \label{ru1}
\end{figure*}

\section{Model and solution}

As shown in Figs.~\ref{ru1} (a-c), we consider a cigar-shaped BEC (approximately $N=1.2 \times 10^5$ ultracold Rb atoms, with mass $m_a$ and atomic transition frequency $\omega_a$) trapped optically inside a high-finesse Fabry-Perot cavity, with the trapping direction parallel to the cavity axis $x$. For a large detuning $\Delta_{a}=\omega_{c}-\omega_{a}$ between the cavity field ($\omega_c$) and the atomic transition frequencies ($\omega_a$), the excited state of the atoms can be adiabatically eliminated and atomic spontaneous emission can thus be neglected~\cite{Morsch06Dynamics}. In this situation, the Hamiltonian of the system can be given by~\cite{Murch08Observation,Brennecke08Cavity,Smith11Optomechanical,Spethmann16Cavity,Purdy10Tunable}
\begin{align}\label{Ham}
H=&\hbar\omega_{c}c^{\dagger}c+\int_{-\frac{L}{2}}^{\frac{L}{2}}dx\Psi^{\dagger}(x)	\bigg{[}\frac{-\hbar^{2}}{2m_{b}}\frac{d^{2}}{dx^{2}}+\hbar U_{0}\mathrm{cos}^{2}(kx)c^{\dagger}c\nonumber\\&+\frac{1}{2}U_{s}\Psi^{\dagger}(x)\Psi(x)\bigg{]}\Psi(x),	
\end{align}
where $L$ is the cavity length; $c$ and $\Psi(x)$ are the annihilation operators of the cavity field and the atomic field, respectively; $k$ is wave number of the cavity field; $U_0=g_0^2/\Delta_a$ is the optical lattice barrier height per photon, where $g_0$ is the coupling strength between single atom and photon; $U_s = 4\pi\hbar^2a_s/m_a$ describes the strength of the interaction between two atoms, where $a_s$ is the two-body $s$-wave scattering length, which can vary from positive to negative values through a broad Feshbach resonance~\cite{Inouye98Observation}.

In the weak-excitation condition, i.e., $U_0\langle a^\dagger a\rangle \leq 10 \omega_R$, where $\omega_R=\hbar k^2/2m_a$ is the recoil frequency of the condensate atoms, the condensate field operator $\Psi(x)$ can be restricted as a sum of the zero-momentum mode and its first two side modes~\cite{Brennecke08Cavity}
\begin{align}\label{Qum}
\Psi(x)=\sqrt{\frac{N}{L}}+\sqrt{\frac{2}{L}}\mathrm{cos}(2kx)b.
\end{align}
Here, the operator $b$ is the annihilation operator of the collective mechanical oscillator of the condensate atoms, corresponding to a small quantum fluctuation around the classical average of the condensate mode ($\sqrt{N/L}$) of the atomic field.

%%%%%%%%%%%%%%%%%%%%%%%%%%%%%%%%%%%%%%%%%%%%%
\begin{figure*}[t]
\centerline{\includegraphics[width=0.98\textwidth]{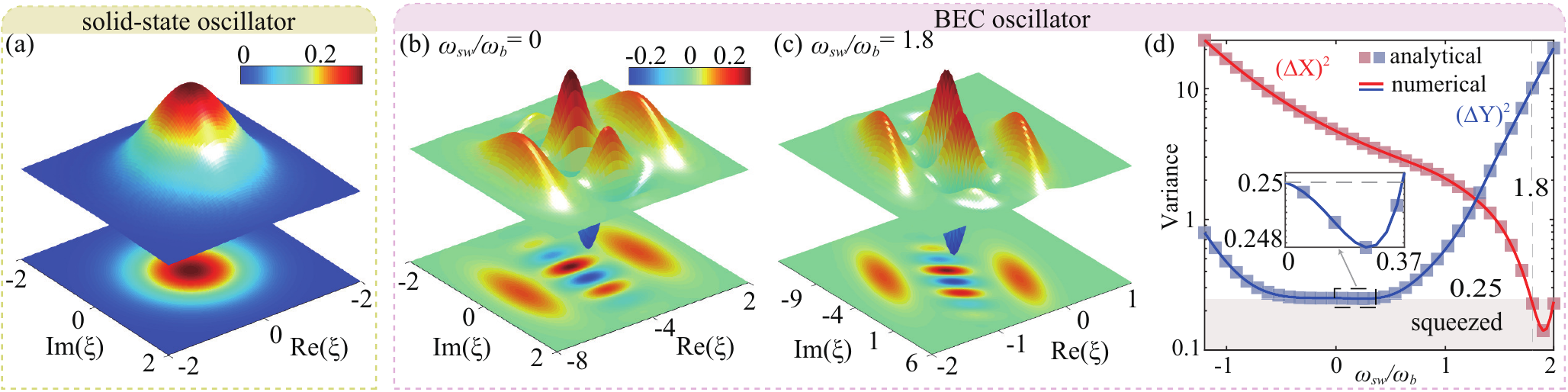}}
\caption{(a-c) Wigner functions of the states for the mechanical mode in a solid-state COM system, and a BEC-based COM system for the $s$-wave scattering frequencies: $\omega_{sw}/\omega_{b}=0, 1.8$. (d) Variances of the position and the momentum in the BEC mechanical mode as a function of the $s$-wave scattering frequency $\omega_{sw}$. The parameters used here are the same as in Fig.~\ref{ru1}.} \label{ru2}
\end{figure*}
%%%%%%%%%%%%%%%%%%%%%%%%%%%%%%%%%%%%%%%%%%%%%%%%%%%
By substituting the wave-function operator in Eq.~(\ref{Qum}) to Eq.~(\ref{Ham}), the Hamiltonian can be rewritten as
\begin{align}\label{Qumham}
H_{1}=&\hbar\omega_{c'}c^{\dagger}c+\hbar\omega_{b}b^{\dagger}b+\frac{1}{4}\hbar\omega_{sw}(b^{2}+b^{\dagger^{2}})+\frac{\hbar}{\sqrt{2}}gc^{\dagger}c(b^{\dagger}+b),
\end{align}
where $\omega_{c'}=\omega_{c}+{NU_{0}}/{2}$ is the effective resonant frequency in the presence of the BEC;  $\omega_{b}$ is the frequency of the mechanical oscillator comprised of a trapped BEC; $\omega_{sw}=8\pi\hbar a_{s}N/m_{b}L\omega^{2}$ is the $s$-wave scattering frequency originating from the atomic collisions, and $\omega$ denotes the waist radius of the optical mode; $g=\sqrt{N}U_{0}/2$ is the COM coupling strength between the optical field and the collective mechanical oscillator of BEC.

In the remainder of this paper, we study how to generate Schr\"{o}dinger cat states in the BEC-based COM system. Based on the Hamiltonian in Eq.~\ref{Qumham}, the time evolution operator for the system is given by ($\hbar=1$)
\begin{align}\label{tiev}
U(t)=&S(r_s)S(-r_se^{-2i\omega_{s}t})e^{-i\omega_{c'}c^{\dagger}ct}e^{i(c^{\dagger}c)^{2}[\delta t-\eta^{2}\mathrm{sin}(\omega_{s}t)]}\nonumber\\&\times e^{ c^{\dagger}c[b^{\dagger}\alpha(t)-b\alpha(t)^{*}]}e^{-i\omega_{s}b^{\dagger}bt},
\end{align}
with
\begin{align}\label{eff}
	r_s =&\frac{1}{4} \mathrm{ln}\frac{2\omega_b+\omega_{sw}}{2\omega_b-\omega_{sw}},\, \omega_{s} = (\omega_b-\frac{1}{2}\omega_{sw})e^{2r_s},
	\nonumber\\
	g_s =&\frac{1}{\sqrt{2}}ge^{-r_s}, \, \eta=-\frac{g_s}{\omega_s}, \, \delta=\frac{g_s^2}{\omega_s}, \, f(t) = \eta(1-e^{-i\omega_st}),
	\nonumber\\ \alpha(t) =& f(t)\mathrm{cosh}(r_s)-f(t)^*e^{-2i\omega_st}\mathrm{sinh}(r_s).
\end{align}

\section{Probabilistic generation of squeezed BEC mechanical Schr\"{o}dinger cat states}
Let us assume that initially the cavity field is in the equal superposition between the vacuum and single-photon states, and the mechanical mode is in the vacuum state, i.e.,
\begin{align}
	|\psi(0)\rangle=\frac{1}{\sqrt{2}}(|0\rangle_{c}+|1\rangle_{c})|0\rangle_{b}.
\end{align}
By using Eq.~(\ref{tiev}), the state of the system in the interaction picture (omitting the free evolution of the cavity field) at time $t$ is obtained as
\begin{align} \label{psi}
|\psi(t)\rangle=&\frac{1}{\sqrt{2}}S(r_s)S(-r_se^{-2i\omega_{s}t})\{|0\rangle_{c}|0\rangle_{b}+e^{i\epsilon(t)}|1\rangle_{c}D_{b}[\alpha(t)]|0\rangle_{b}\},
\end{align}
with $\epsilon(t)=\delta t-\eta^{2}\mathrm{sin}(\omega_{s}t)$. Equation~(\ref{psi}) describes an entangled state involving the cavity mode and the BEC mechanical resonator.

We take the concurrence $\mathcal{C}$ as a measure of entanglement via the method in Ref.~\cite{Kuang03Generation}. The concurrence varies from $\mathcal{C}=0$ for a separable state to $\mathcal{C}=1$ for a maximally entangled state. In Fig.~\ref{ru1}(f), we show the dynamics of the concurrence $\mathcal{C}$ for the state $|\psi(t)\rangle$ with the $s$-wave scattering frequency $\omega_{sw}/\omega_{b} =1$. We find that the concurrence $\mathcal{C}$  exhibits an oscillating behavior with a period of $T=2\pi/\omega_{s}$, which originates from the time factor $e^{-i\omega_{s}t}$ in $f(t)$. At time $t=2n\pi/\omega_s$ for a natural number $n$, we have concurrence $\mathcal{C}=0$, which means that the entanglement disappears. This phenomenon can be understood as follows: the coherent amplitude of $\alpha(t)=0$ is achieved with $f(t)=0$ at this moment, and the state of the system becomes the direct product states of the cavity mode and the BEC mechanical oscillator again. In the middle duration of a period, the concurrence $\mathcal{C}$ reachs its maximum value, i.e., $\mathcal{C}=1$.

After a projective measurement of the cavity field with a projector $P=\sum_{\nu=\pm}|{\phi_{\nu}}\rangle_{c}\langle{\phi_{\nu}}|$, where $|\phi_\pm\rangle_c=(|0\rangle_{c}\pm|1\rangle_{c})/\sqrt{2}$, the BEC mechanical mode can be prepared in a squeezed cat state
\begin{align}\label{BECcat}	|\psi_\pm(t)\rangle_b=&\frac{1}{2}\mathcal{N}_{\pm}S(r_s)S(-r_se^{-2i\omega_{s}t})\{|0\rangle_{b}
	\pm e^{i\epsilon(t)}D_{b}[\alpha(t)]|0\rangle_{b}\}.
\end{align}
where $\mathcal{N}_{\pm}=2\{2\pm2e^{-|\alpha(t)|^2/2}\mathrm{Re}[e^{i\epsilon(t)}]\}^{-1/2}$are the normalization constants. When the $s$-wave scattering frequency $\omega_{sw}=0$, we have the maximum coherent amplitude $|\alpha|_\mathrm{max}=\sqrt{2} g/\omega_b$. In Fig.~\ref{ru1}(d), we plot the coherent amplitude $|\alpha(t)|$ as a function of the scaled time $\omega_{b} t$ for $\omega_{sw}=0$. We have $|\alpha|_\mathrm{max}\approx4.2$ with the parameters $\omega_{b}= 2\times10^5~\mathrm{Hz}$~\cite{Spethmann16Cavity} and $g=6\times10^5~\mathrm{Hz}$~\cite{Murch08Observation} in the simulations. The maximal amplitude of the generated coherent states in the BEC-based COM system is about $3$ order of magnitude larger than that ($\sim10^{-3}$) obtained in the solide-state COM systems~\cite{Verhagen12Quantum}, which are composed of a high-Q optical cavity and, e.g., a $\mathrm{SiO_2}$ mechanical oscillator, with an experimentally feasible COM coupling strength $g\sim10^4\mathrm{Hz}$, and a mechanical frequency $\omega_b\sim10^7\mathrm{Hz}$.
The large coherent amplitude obtained in the BEC-based COM system, thus with a small overlap between the two coherent states $\langle-\alpha|\alpha\rangle\ll1$, has important applications in quantum information processing.
Moreover, the coherent amplitude can be enlarged further by modulating the $s$-wave scattering in the BEC-based COM system. In Fig.~\ref{ru1}(e), we show the amplitude $|\alpha|$ versus the $s$-wave scattering frequency $\omega_{sw}$. We find that the coherent amplitude can be enlarged by a factor of $5$ for the $s$-wave scattering frequency $\omega_{sw} \to -2\omega_b$. This is potentially beneficial to different scenarios of quantum metrology.

%%%%%%%%%%%%%%%%%%%%%%%%%%%%%%%%%%%%%%%%%%%%%%%%%%%%%
\begin{figure}[t]
	\centerline{\includegraphics[width=0.45\textwidth]{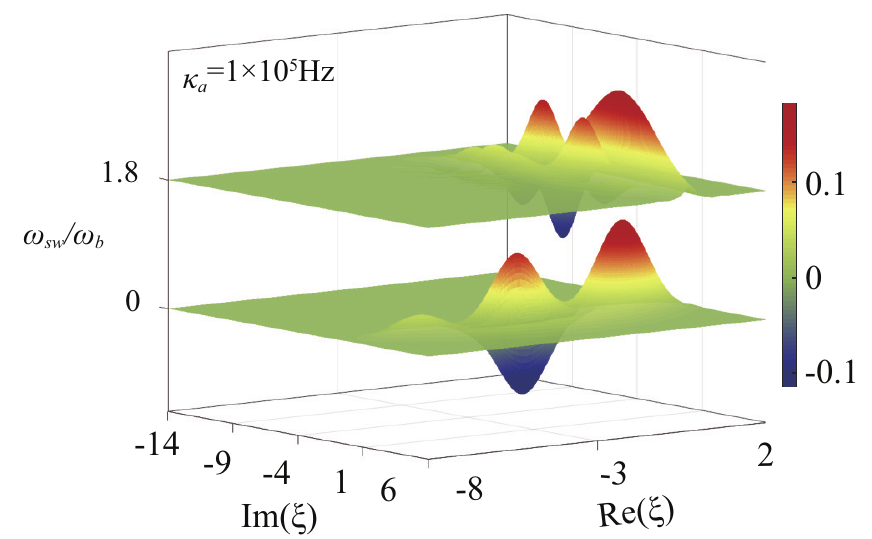}}
	\caption{The Wigner functions of the state $|\psi_+(t)\rangle_b$ at time $t=\pi/\omega_b$ with the cavity loss $\kappa_a=10^5 \mathrm{Hz}$ for the $s$-wave scattering frequencies: $\omega_{sw}/\omega_{b}=0, 1.8$. The other parameters used here are the same as in Fig.~\ref{ru1}.} \label{ru4}
\end{figure}
%%%%%%%%%%%%%%%%%%%%%%%%%%%%%%%%%%%%%%%%%%%%%%%%%%%%%

In order to show quantum interference and coherence in the generated states intuitively, we examine the Wigner function of the mechanical mode~\cite{Walls94Quantum}. The Wigner function of the BEC mechanical mode is defined by
\begin{align}\label{win}
W(\xi)=\frac{2}{\pi}\mathrm{Tr}[D^\dagger(\xi)\rho_{b} D(\xi)(-1)^{b^\dagger b}],
\end{align}
where $\rho_{b}$ is the density matrix of the BEC mechanical mode, and $D(\xi)=\mathrm{exp}(\xi b^\dagger-\xi^*b)$ describes a displacement operator with a complex variable $\xi$. In Figs.~\ref{ru2}(a) and \ref{ru2}(b), we plot the Wigner functions for the mechanical mode at $t=\pi/\omega_b$ for $\omega_{sw}=0$ in the solid-state~\cite{Verhagen12Quantum}, and the BEC-based COM systems, respectively. We find that there is only one peak of the Wigner function in phase space for the solid-state COM system. That is because the COM strength in the solid-state COM system is too weak to distinguish the two coherent states in phase space ($|\alpha\rangle$ and $|-\alpha\rangle$), i.e., $\langle-\alpha|\alpha\rangle\approx1$. In sharp contrast, in the BEC-based COM system, we can see two main peaks of the Wigner function in phase space, which represent a vacuum state and a coherent state, respectively. The Wigner function also exhibits a distinct oscillating pattern between these two peaks. This oscillating feature is a clear evidence of quantum interference, and macroscopically distinct superposition components. Furthermore, we find that the distance of the two peaks in phase space can be modulated by appropriately tuning the $s$-wave scattering length, shown in Fig.~\ref{ru2}(c).

Now we study the role of collisions on the squeezing in the generated cat states. Let us consider the position and momentum operators of the BEC mechanical oscillator $X=(b^\dagger + b)/2$ and $Y=i(b^\dagger - b)/2$. We analyze the mechanical squeezing by analytically and numerically evaluating the variances of the position and the momentum, i.e., $\left(\Delta X\right)^{2}$ and $\left(\Delta Y\right)^{2}$. If either $(\Delta X)^2$ or $(\Delta Y)^2$ is below $0.25$, the state of the BEC mechanical oscillator then exhibits quadrature squeezing~\cite{Walls94Quantum}. For the state $|\psi_+(t)\rangle_b$, we obtain
\begin{align}\label{squeX}
(\Delta X)^2=&_b\langle\psi_+(t)|X^2|\psi_+(t)\rangle_b-(_b\langle\psi_+(t)|X|\psi_+(t)\rangle_b)^2\\ \nonumber =&\frac{\mathcal{N}_{+}^2}{8}e^{-2r_s}\{s(t)+\mathrm{Re}[h(t)\alpha(t)^2]+s(t)|\alpha(t)|^2\} \\ \nonumber &-\frac{\mathcal{N}_{+}^4}{32}e^{-2r_s} \{\mathrm{Re}[g(t)^2\alpha(t)^2] + |g(t)|^2|\alpha(t)|^2 \}
\\ \nonumber &+\frac{\mathcal{N}_{+}^2}{8}e^{-2r_s}e^{-\frac{|\alpha(t)|^2}{2}}\{\mathrm{Re}[s(t)e^{i\epsilon(t)}]+\mathrm{Re}[h(t)\alpha(t)^2e^{i\epsilon(t)}]\} \\ \nonumber
&-\frac{\mathcal{N}_{+}^4}{32}e^{-2r_s}e^{-|\alpha(t)|^2}\{\mathrm{Re}[g(t)^2\alpha(t)^2e^{2i\epsilon(t)}]+|g(t)|^2|\alpha(t)|^2\} \\ \nonumber
&-\frac{\mathcal{N}_{+}^4}{16}e^{-2r_s}e^{-\frac{|\alpha(t)|^2}{2}}\{\mathrm{Re}[|g(t)|^2|\alpha(t)|^2e^{i\epsilon(t)}] \\ \nonumber
&+\mathrm{Re}[g(t)^2\alpha(t)^2e^{i\epsilon(t)}]\}
,
\end{align}
and
\begin{align}\label{squeY}
(\Delta Y)^2=&_b\langle\psi_+(t)|Y^2|\psi_+(t)\rangle_b-(_b\langle\psi_+(t)|Y|\psi_+(t)\rangle_b)^2\\ \nonumber =&\frac{\mathcal{N}_{+}^2}{8}e^{2r_s}\{s(t)-\mathrm{Re}[h_1(t)\alpha(t)^2]+s_1(t)|\alpha(t)|^2\} \\ \nonumber &+\frac{\mathcal{N}_{+}^4}{32}e^{2r_s}\{\mathrm{Re}[g_1(t)^2\alpha(t)^2] -|g_1(t)|^2|\alpha(t)|^2 \}
\\ \nonumber &+\frac{\mathcal{N}_{+}^2}{8}e^{2r_s}e^{-\frac{|\alpha(t)|^2}{2}}\{\mathrm{Re}[s_1(t)e^{i\epsilon(t)}]-\mathrm{Re}[h_1(t)\alpha(t)^2e^{i\epsilon(t)}]\} \\ \nonumber
 &+\frac{\mathcal{N}_{+}^4}{32}e^{2r_s}e^{-|\alpha(t)|^2}\{\mathrm{Re}[g_1(t)^2\alpha(t)^2e^{2i\epsilon(t)}]-|g_1(t)|^2|\alpha(t)|^2\} \\ \nonumber
  &+\frac{\mathcal{N}_{+}^4}{16}e^{2r_s}e^{-\frac{|\alpha(t)|^2}{2}}\{\mathrm{Re}[g_1(t)^2\alpha(t)^2e^{i\epsilon(t)}]\\ \nonumber
  &-\mathrm{Re}[|g_1(t)|^2|\alpha(t)|^2e^{i\epsilon(t)}]\},
\end{align}
where
\begin{align}
g(t) =& \mathrm{cosh}(r_s)+e^{2i\omega_st}\mathrm{sinh}(r_s),\\ \nonumber
g_1(t) =& \mathrm{cosh}(r_s)-e^{2i\omega_st}\mathrm{sinh}(r_s),\\ \nonumber
h(t)=&\mathrm{cosh}^2(r_s)+ e^{2i\omega_st}\mathrm{sinh}(2r_s)+e^{4i\omega_st}\mathrm{sinh}^2(r_s),
\\ \nonumber
h_1(t)=&\mathrm{cosh}^2(r_s)- e^{2i\omega_st}\mathrm{sinh}(2r_s)+e^{4i\omega_st}\mathrm{sinh}^2(r_s),
\\ \nonumber
s(t)=&\frac{1}{2}\mathrm{sinh}(2r_s)(e^{-2i\omega_st}+e^{2i\omega_st})+\mathrm{cosh}(2r_s),\\ \nonumber
s_1(t)=&-\frac{1}{2}\mathrm{sinh}(2r_s)(e^{-2i\omega_st}+e^{2i\omega_st})+\mathrm{cosh}(2r_s).
\end{align}

The variances  $(\Delta X)^2$ and $(\Delta Y)^2$ are ploted as functions of the $s$-wave scattering frequency $\omega_{sw}$ in Fig.~\ref{ru2}(d).  We find that $(\Delta X)^2$ decreases monotonously, and $(\Delta Y)^2$ decreases first and then increases with the increase of the $s$-wave scattering frequency.  According to the values of the variances  $(\Delta X)^2$ and $(\Delta Y)^2$ at time $t=\pi/\omega_b$, the value of the scattering frequency $\omega_{sw}$ can be divided into three areas. (i) When  $-2<\omega_{sw}/\omega_{b}<0$ and $0.37<\omega_{sw}/\omega_{b}<1.78$, both the variances $(\Delta X)^2$ and $(\Delta Y)^2$ are always larger than $0.25$, which means the squeezing disappears in both directions. (ii) When $0<\omega_{sw}/\omega_{b}<0.37$, the $(\Delta X)^2$ is larger than $0.25$, but $(\Delta Y)^2$ is less than $0.25$, representing that the squeezing occurs in the momentum quadrature. (iii) When $1.78<\omega_{sw}/\omega_{b}<2$, the $(\Delta X)^2$ is always less than $0.25$, but $(\Delta Y)^2$ is larger than $0.25$, representing that the squeezing occurs in the position quadrature. Hence, a squeezed BEC mechanical cat state is generated, and the squeezing direction of the cat state can be switched by modulating the $s$-wave scattering frequency $\omega_{sw}$ in the BEC-based COM system.

In order to study the effect of the cavity loss on the state generation, we numerically integrate the master equation~\cite{Walls94Quantum},
\begin{align}\label{master}
\dot{\rho}=&-i[H_1,\rho]+\kappa_a\mathcal{L}[a]\rho,
\end{align}
where $\mathcal{L}[o]=2o\rho o^\dagger-o^\dagger o\rho - \rho o^\dagger o$ is the standard Lindblad superoperator; $\kappa_a$ is the photon loss rate of the cavity field. It is worth noting that here we only consider the effect of the optical cavity decay. The reasons for this are given as follows. (i) The decay rate of the BEC mechanical oscillator in the BEC-based COM system is low ($\sim10^3\mathrm{Hz}$~\cite{Murch08Observation,Brennecke08Cavity,Smith11Optomechanical,Spethmann16Cavity,Purdy10Tunable}), so the time scale of the dissipative effect (ms) is much larger than that of the generation of cat states and also the cavity photon lifetime ($0.1\mu$s).
(ii) The atoms in BEC can be evaporatively cooled to around $\mu$K-nK~\cite{Murch08Observation,Brennecke08Cavity,Smith11Optomechanical,Spethmann16Cavity,Purdy10Tunable}, thus the mean thermal phonon
number in this mechanical mode is smaller than $0.01$, which can be ignored. In Fig.~\ref{ru4}, we plot the Wigner function of the state $|\psi_+(t)\rangle_{b}$ with the cavity loss rate $\kappa_a = 1\times10^5\mathrm{Hz}$. We can see that the interference pattern and the peaks of the Wigner function are attenuated gradually due to dissipation of the optical cavity, which means that dissipation destroys the macroscopic quantum coherence.
%%%%%%%%%%%%%%%%%%%%%%%%%%%%%%%%%%%%%%%%%%%%%
\begin{figure*}[t]
	\centerline{\includegraphics[width=0.95\textwidth]{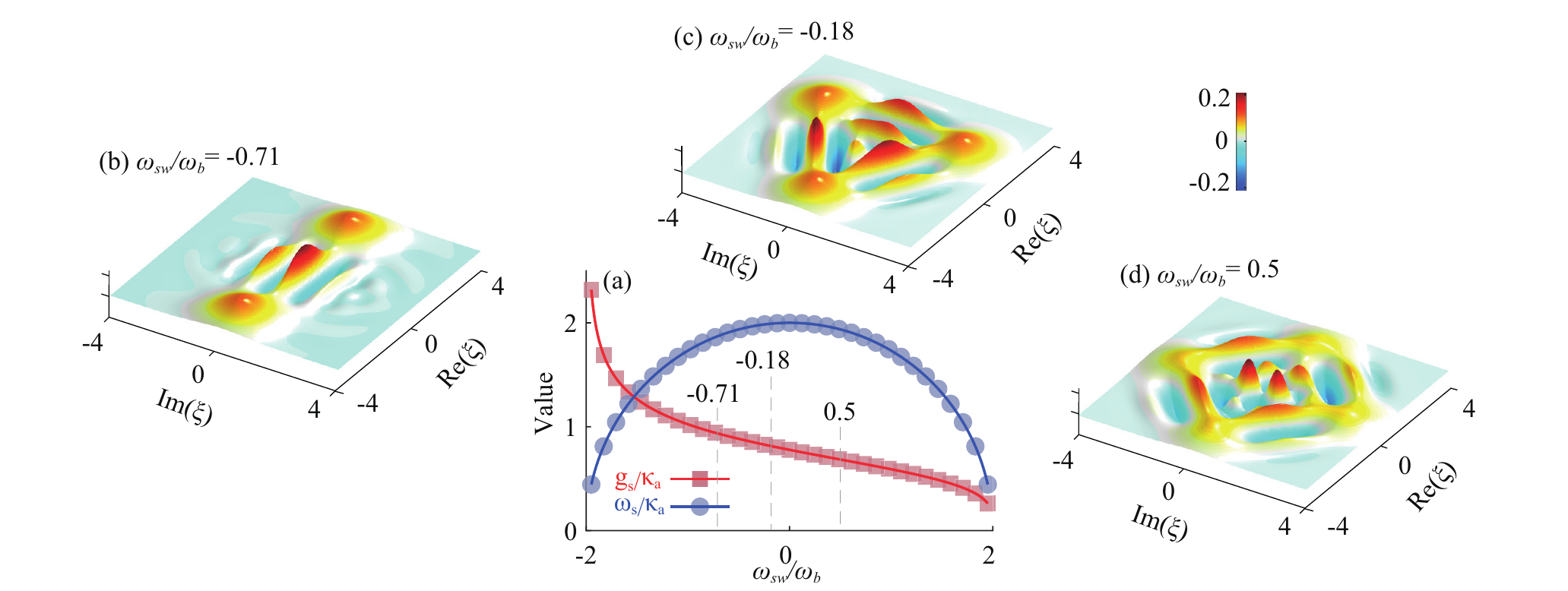}}
	\caption{(a) Effective COM coupling strength $g_s$, and the BEC mechanical mode frequency $\omega_s$ versus the s-wave scattering frequency $\omega_{sw}$. (b-d) The Wigner functions of the optical mode by tracing over the BEC degrees of freedom at a time $t=2\pi/\omega_s$ with different $s$-wave scattering frequencies: $\omega_{sw}/\omega_{b}=-0.71, -0.18, 0.5$. Here, we assume that $g=1.1\times10^5~\mathrm{Hz}$, and $\kappa_a=1\times10^5~\mathrm{Hz}$, and the other parameters are the same as in Fig.~\ref{ru1}.} \label{ru3}
\end{figure*}
%%%%%%%%%%%%%%%%%%%%%%%%%%%%%%%%%%%%%%%%%%%%%%%%%%%
\section{Deterministic generation of optical multicomponent Schr\"{o}dinger cat states}
In the BEC-based COM system, the $s$-wave scattering frequency $\omega_{sw}$ of the atomic collisions is experimentally controllable through the frequency of the BEC transverse trap~\cite{Inouye98Observation}. Thus, this gives an extra degree of freedom to control the optical Schr\"{o}dinger cat states generated in the cavity field. As shown in Fig.~\ref{ru3}(a), we find that the effective COM coupling strength $g_s$ and the mechanical mode frequency $\omega_s$ change synchronously with the scattering frequency $\omega_{sw}$. Let us assume that initially the cavity field is in the coherent state of $\alpha=2$ and the mechanical mode is in the vacuum state, i.e.,
\begin{align}\label{iniop}
|\varphi(0)\rangle=|\alpha\rangle_{c}|0\rangle_{b},
\end{align}
By using Eq.~(\ref{tiev}), the state of the system in the interaction picture at time $t$ is given by
\begin{align} \label{phi}
|\varphi(t)\rangle=&e^{-|\alpha|^2/2}\sum_{n=0}^{\infty}\frac{\alpha^n}{\sqrt{n!}}e^{in^{2}[\delta t-\eta^{2}\mathrm{sin}(\omega_{s}t)]}|n\rangle_{c}\nonumber\\&\otimes S(r_s)S(-r_se^{-2i\omega_{s}t})|n\alpha(t)\rangle_{b},
\end{align}
where $|n\rangle_{c}$ represents the Fock state of the cavity mode with the number of photons $n$. When we consider a special time, i.e., $\tau=2\pi/\omega_{s}$, the BEC mechanical oscillator returns to its original state, i.e., $S(r_s)S(-r_se^{-2i\omega_{s}\tau})|n\alpha(\tau)\rangle_{b}=|0\rangle_{b}$. The state of the system becomes the original direct product state of the cavity mode and the BEC mechanical oscillator. In this way, the state of the cavity field can be obtained by tracing over the BEC degree of freedom,
\begin{align} \label{phic}
|\varphi(\tau)\rangle_c=&e^{-|\alpha|^2/2}\sum_{n=0}^{\infty}\frac{\alpha^n}{\sqrt{n!}}e^{in^{2}2\pi{g_s^2}/{\omega_s^2} }|n\rangle_{c}.
\end{align}
According to the same method of generating the optical Schr\"{o}dinger cat states discussed in Ref.~\cite{Bose97Preparation}, we can achieve multicomponent optical Schr\"{o}dinger cat states with various $s$-wave scattering frequencies $\omega_{sw}$. For $\omega_{sw}=-0.71\omega_b$, i.e., $({g_s}/{\omega_s})^2\approx1/4$, this state $|\varphi(\tau)\rangle_c$ can be converted into an optical two-component Schr\"{o}dinger cat state
\begin{align} \label{phi2c}
|\varphi(\tau)\rangle_c\approx e^{-\frac{|\alpha|^2}{2}}\left[\left(\frac{1+i}{2}\right)|\alpha\rangle_c+\left(\frac{1-i}{2}\right)|-\alpha\rangle_c\right].
\end{align}
For $\omega_{sw}=-0.18\omega_b$, i.e., $({g_s}/{\omega_s})^2\approx1/6$, the state $|\varphi(\tau)\rangle_c$ becomes a three-component Schr\"{o}dinger cat state
\begin{align} \label{phi3c}
|\varphi(\tau)\rangle_c\approx c_1|-\alpha\rangle_c+c_2|\alpha e^{i\pi/3}\rangle_c+c_3|\alpha e^{-i\pi/3}\rangle_c,
\end{align}
where $c_1=-i\mathrm{sin(\pi/3)}/\left[1+\mathrm{cos(\pi/3)}\right]$ and $c_2=c_3=(1+e^{i(\pi/3)})/2\left[1+\mathrm{cos(\pi/3)}\right]$. For $\omega_{sw}=0.5\omega_b$, i.e., (${g_s}/{\omega_s})^2\approx1/8$, the state $|\varphi(\tau)\rangle_c$ can be turned into a four-component Schr\"{o}dinger cat state
\begin{align} \label{phi4c}
|\varphi(\tau)\rangle_c\approx \frac{e^{i\pi/4}}{2}(|\alpha\rangle_c-|-\alpha\rangle_c)+\frac{1}{2}(|i\alpha\rangle_c+|-i\alpha\rangle_c).
\end{align}

In Figs.~\ref{ru3}(b-d), we numerically plot the Wigner function of the optical mode by tracing over the BEC degree of freedom at a time $t=2\pi/\omega_s$ with various $s$-wave scattering frequencies $\omega_s$. We find that the numerical results are in good agreement with the analytical ones. When we consider a $s$-wave scattering frequency $\omega_{sw}/\omega_b=-0.71$, an optical two-component cat state appears, whereas when $\omega_{sw}/\omega_b=-0.18$ and $0.5$, three-component, and four-component cat states appear, respectively.  It is worth noting that the multicomponent optical Schr\"{o}dinger cat states are also proposed in the COM system~\cite{Bose97Preparation} by tunning the rate of the COM coupling strength $g$ and the frequency of the mechanical oscillator $\omega_b$, but it is a big challenge to control these parameters in an established system.

In general, to observe these generated cat states, the COM systems need to work in the sideband-resolved regime, i.e., $\omega_{b}>\kappa_{a}$. We also notice that in current experiments, the BEC-based COM system is often operated in the
unresolved-sideband regime~\cite{Murch08Observation,Brennecke08Cavity,Smith11Optomechanical,Spethmann16Cavity,Purdy10Tunable}, i.e., $\omega_{b}/\kappa_{a}<1$. This problem may be circumvented by using two schemes: (i) moderately increasing the BEC trapping frequency to increase the mechanical mode resonance frequency $\omega_{b}$; (ii) properly increasing the dissipation time of the cavity. The dissipation time can be extended by two ways. On the one hand, we can improve the $Q$ factor of the optical cavity through the advancement of process technology. We find that the condition
$\omega_{b}/\kappa_{a}\sim0.5$ has been implemented in BEC-based COM systems~\cite{Smith11Optomechanical}. In the near future, the $Q$ factor of the optical resonator can be further improved, allowing one to reach the resolved-sideband regime. For instance, Savchenkov \textit{et al}., have reported some $Q$ factors as high as $10^{11}$ with calcium fluoride resonators~\cite{Savchenkov07Optical}, and by using a chemo-mechanical polishing procedure, a silica microdisk with ultra-high quality factors ($>10^8$) can be fabricated~\cite{Honari21Fabrication}. On the other hand,
we can replace the disordered atomic cloud with an ordered two-dimensional (2D) array of atoms, so as to reduce loss caused by transverse scattering from the atoms~\cite{Shahmoon20Cavity}. In addition, we can increase the robustness against the decay of the system by rapidly generating large-size Schr\"{o}dinger cat states. Recently, Chen \textit{et al}., have shown that the shortcuts-to-adiabatic protocol can generate nonclassical states, with a preparation process $\sim10$ times faster compared to adiabatic protocols in the Rabi model via parametric amplification~\cite{Chen21Shortcuts,Qin18Exponentially}.

\section{Conclusions} \label{C}
In summary, by using the unique properties of BEC-based COM systems, we have shown how to create cat states of mechanical vibrations with distinguishable superposition components, how to make use of this cat state to generate and manipulate mechanical quadrature squeezing, and how to achieve multicomponent optical cat states as well, which are otherwise difficult to be attainable in solid-state COM systems. These findings open up a new way to explore the nonclassical nature of massive objects by utilizing diverse BEC gases such as annular BEC~\cite{Kumar21Cavity}, spinor BEC~\cite{Jing11Quantum,Stenger98Spin}, multi-component BEC~\cite{Myatt97Production,Kuang07Generation} and molecular Fermi-BEC~\cite{Greiner03Emergence}, in which photons and phonons would experience different types of light-motion interaction. Moreover, in a broader view, the ability to reach the ultrastrong light-motion coupling regime would also make the BEC-based COM systems a promising platform to engineer a variety of nonlinear COM effects, such as solitons, optical multistable oscillating~\cite{Dong11Multistability}, and phonon lasing.

~\\	
%%%%%%%%%%%%%%%%%%%%%%%%%%%%%%%%%%%%%%%%%%%%%%%%%%%%%%%
%%% Acknowledgements.
%%%%%%%%%%%%%%%%%%%%%%%%%%%%%%%%%%%%%%%%%%%%%%%%%%%%%%%
\textbf{Declaration of Competing Interest}

The authors declare that they have no conflicts of interest in this
work.
~\\	
\textbf{Acknowledgements}

We thank Jie-Qiao Liao and YunLan Zuo for helpful discussions. H.J. was supported by the National Natural Science Foundation of China (NSFC) (Grants No. 11935006 and No. 11774086) and the Science and Technology Innovation Program of Hunan Province (Grant No. 2020RC4047). L.-M.K. was supported by the NSFC (Grants No. 1217050862, 11935006 and No. 11775075). X.-W.X. was supported by the NSFC (Grants No. 12064010), and Natural Science Foundation of Hunan Province of China (Grant No. 2021JJ20036). Y.-F.J. was supported by the NSFC (Grant No. 12147156), the China Postdoctoral Science Foundation (Grant No. 2021M701176) and the Science and Technology Innovation Program of Hunan Province (Grant No. 2021RC2078). B.J.L. was supported by Postgraduate Scientific Research  Innovation Project of Hunan Province (Grant No. CX20210471).

%%%%%%%%%%%%%%%%%%%%%%%%%%%%%%%%%%%%%%%%%%%%%%%%%%%%%%%
%%% Appendix sections. ??????, ????
%%%%%%%%%%%%%%%%%%%%%%%%%%%%%%%%%%%%%%%%%%%%%%%%%%%%%%%


\section*{References}

\begin{thebibliography}{99}
	
\bibitem{Zurek03Decoherence}
W.~H.~Zurek, Decoherence, einselection, and the quantum origins of the classical, Rev. Mod. Phys. 75 (2003) 715--775; https://doi.org/10.1103/RevModPhys.75.715.
S.~Haroche, Nobel lecture: Controlling photons in a box and exploring the quantum to classical boundary, ibid. 85 (2013) 1083--1102. https://doi.org/10.1103/RevModPhys.85.1083.		

\bibitem{Ku20Experimental}
H.-Y.~Ku, N.~Lambert, F.-J.~Chan, C.~Emary, Y.-N.~Chen, and F.~Nori, Experimental test of non-macrorealistic cat states in the cloud, npj Quantum Information 6 (2020) 98; https://doi.org/10.1038/s41534-020-00321-x.
W.~Cai, Y.~Ma, W.~Wang, C.-L.~Zou, and L.~Sun, Bosonic quantum error correction codes in superconducting quantum circuits, Fund. Res. 1 (2021) 50--67. https://doi.org/10.1016/j.fmre.2020.12.006.

\bibitem{Pezze18Quantum}	
L.~Pezz\`{e}, A.~Smerzi, M.~K.~Oberthaler, R.~Schmied, and P.~Treutlein, Quantum metrology with nonclassical states of atomic ensembles, Rev. Mod. Phys. 90 (2018) 035005. https://doi.org/10.1103/RevModPhys.90.035005.

\bibitem{Ourjoumtsev07Generation}
A.~Ourjoumtsev, H.~Jeong, R.~Tualle-Brouri, and P.~Grangier, Generation of optical 'Schr\"{o}dinger cats' from photon  number states, Nature (London) 448 (2007) 784--786. https://doi.org/10.1038/nature06054.

\bibitem{Ourjoumtsev06Generating}
A.~Ourjoumtsev, R.~Tualle-Brouri, J.~Laurat, and P.~Grangier, Generating Optical Schrdinger Kittens for Quantum Information Processing, Science 312 (2006) 83--86. https://doi.org/10.1126/science.1122858.

\bibitem{Nielsen06Generation}%%%%%%%%%%%%%%%%%%%%%%%%%%%%%%%%%%%%%%%%%%%%%%%%%%%
J.~S.~Neergaard-Nielsen, B.~M.~Nielsen, C.~Hettich, K.~M{\o}lmer, and E.~S.~Polzik, Generation of a Superposition of Odd Photon Number States for Quantum Information Networks, Phys. Rev. Lett. 97 (2006) 083604. https://doi.org/10.1103/PhysRevLett.97.083604.

\bibitem{Vlastakis13Deterministically}
B.~Vlastakis, G.~Kirchmair, Z.~Leghtas, S.~E.~Nigg, L.~Frunzio, S.~M.~Girvin, M.~Mirrahimi, M.~H.~Devoret, and R.~J.~Schoelkopf, Deterministically Encoding Quantum Information Using 100-Photon Schr\"{o}dinger Cat States, Science 342 (2013) 607--610. https://doi.org/10.1126/science.1243289.

\bibitem{Kirchmair13Observation}
G.~Kirchmair, B.~Vlastakis, Z.~Leghtas, S.~E.~Nigg, H.~Paik, E.~Ginossar, M.~Mirrahimi, L.~Frunzio, S.~M.~Girvin, and R.~J.~Schoelkopf, Observation of quantum state collapse and revival due to the single-photon Kerr effect, Nature (London) 495 (2013) 205--209. https://doi.org/10.1038/nature11902.

\bibitem{Sychev17Enlargement}
D.~V.~Sychev, A.~E.~Ulanov, A.~A.~Pushkina, M.~W.~Richards, I.~A.~Fedorov, and A.~I.~Lvovsky, Enlargement of optical Schr\"{o}dinger's cat states, Nat. Photon. 11 (2017) 379--382. https://doi.org/10.1038/nphoton.2017.57.

\bibitem{Leghtas15Confining}
Z.~Leghtas, S.~Touzard, I.~M.~Pop, A.~Kou, B.~Vlastakis, A.~Petrenko, K.~M.~Sliwa, A.~Narla, S.~Shankar, M.~J.~Hatridge, M.~Reagor, L.~Frunzio, R.~J.~Schoelkopf, M.~Mirrahimi, and M.~H.~Devoret, Confining the state of light to a quantum manifold by engineered two-photon loss, Science 347 (2015) 853--857. https://doi.org/10.1126/science.aaa2085.

\bibitem{Grimm20Stabilization}
A.~Grimm, N.~E.~Frattini, S.~Puri, S.~O.~Mundhada, S.~Touzard, M.~Mirrahimi, S.~M.~Girvin, S.~Shankar, and M.~H.~Devoret, Stabilization and operation of a Kerr-cat qubit, Nature (London) 584 (2020) 205--209. https://doi.org/10.1038/s41586-020-2587-z.

\bibitem{Monroe96A}
C.~Monroe, D.~M.~Meekhof, B.~E.~King, and D.~J.~Wineland, A "Schr\"{o}dinger Cat" Superposition State of an Atom, Science 272 (1996) 1131--1136. https://doi.org/10.1126/science.272.5265.1131.

\bibitem{Myatt00Decoherence}
C.~J.~Myatt, B.~E.~King, Q.~A.~Turchette, C.~A.~Sackett, D.~Kielpinski, W.~M.~Itano, C.~Monroe, and D.~J.~Wineland, Decoherence of quantum superpositions through coupling to engineered reservoirs, Nature (London) 403 (2000) 269--273. https://doi.org/10.1038/35002001.

\bibitem{Leibfried05Creation}
D.~Leibfried, E.~Knill, S.~Seidelin, J.~Britton, R.~B.~Blakestad, J.~Chiaverini, D.~B.~Hume, W.~M.~Itano, J.~D.~Jost, C.~Langer, R.~Ozeri, R.~Reichle, and D.~J.~Wineland, Creation of a six-atom 'Schr\"{o}dinger cat' state, Nature (London) 438 (2005) 639--642. https://doi.org/10.1038/nature04251.

\bibitem{Omran19Generation}
A.~Omran, H.~Levine, A.~Keesling, G.~Semeghini, T.~T.~Wang, S.~Ebadi, H.~Bernien, A.~S.~Zibrov, H.~Pichler, S.~Choi, J.~Cui, M.~Rossignolo, P.~Rembold, S.~Montangero, T.~Calarco, M.~Endres, M.~Greiner, V.~Vuleti\'{c}, and M.~D.~Lukin, Generation and manipulation of Schr\"{o}dinger cat states in Rydberg atom arrays, Science 365 (2019) 570--574. https://doi.org/10.1126/science.aax9743.

\bibitem{Lu19Global}
Y.~Lu, S.~Zhang, K.~Zhang, W.~Chen, Y.~Shen, J.~Zhang, J.-N.~Zhang, and K.~Kim, Global entangling gates on arbitrary ion qubits, Nature (London) 572 (2019) 363--367; 10.1038/s41586-019-1428-4.
C.~Figgatt, A.~Ostrander, N.~M.~Linke, K.~A.~Landsman, D.~Zhu, D.~Maslov, and C.~Monroe, Parallel entangling operations on a universal ion-trap quantum computer, ibid. 572 (2019) 368--372. https://doi.org/10.1038/s41586-019-1427-5.

\bibitem{Song19Generation}
C.~Song, K.~Xu, H.~Li, Y.-R.~Zhang, X.~Zhang, W.~Liu, Q.~Guo, Z.~Wang, W.~Ren, J.~Hao, H.~Feng, H.~Fan, D.~Zheng, D.-W. ~Wang, H.~Wang, and S.-Y.~Zhu, Generation of multicomponent atomic Schr\"{o}dinger cat states of up to 20 qubits, Science 365 (2019) 574--577. https://doi.org/10.1126/science.aay0600.

\bibitem{Chen21Shortcuts}
Y.-H.~Chen, W.~Qin, X.~Wang, A.~Miranowicz, and F.~Nori, Shortcuts to Adiabaticity for the Quantum Rabi Model: Efficient Generation of Giant Entangled Cat States via Parametric Amplification, Phys. Rev. Lett. 126 (2021) 023602. https://doi.org/10.1103/PhysRevLett.126.023602.

\bibitem{Khazali18Progress}
M.~Khazali, Progress towards macroscopic spin and mechanical superposition via Rydberg interaction, Phys. Rev. A 98 (2018) 043836. https://doi.org/10.1103/PhysRevA.98.043836.

\bibitem{Qin21Generating}
W.~Qin, A.~Miranowicz, H.~Jing, and F.~Nori, Generating Long-Lived Macroscopically Distinct Superposition States in Atomic Ensembles, Phys. Rev. Lett. 127 (2021) 093602. https://doi.org/10.1103/PhysRevLett.127.093602.

\bibitem{Bose97Preparation}
S.~Bose, K.~Jacobs, and P.~L.~Knight, Preparation of nonclassical states in cavities with a moving mirror, Phys. Rev. A 56 (1997) 4175. https://doi.org/10.1103/PhysRevA.56.4175.

\bibitem{Tan13Generation}
H.~Tan, F.~Bariani, G.~Li, and P.~Meystre, Generation of macroscopic quantum superpositions of optomechanical oscillators by dissipation, Phys. Rev. A 88 (2013) 023817; https://doi.org/10.1103/PhysRevA.88.023817.
I.~Shomroni, L.~Qiu, and T.~J.~Kippenberg, Optomechanical generation of a mechanical catlike state by phonon subtraction, ibid. 101 (2020) 033812. https://doi.org/10.1103/PhysRevA.101.033812.

\bibitem{Liao16Macroscopic}
J.-Q.~Liao and L.~Tian, Macroscopic Quantum Superposition in Cavity Optomechanics, Phys. Rev. Lett. 116 (2016) 163602; https://doi.org/10.1103/PhysRevLett.116.163602.
W.~Marshall, C.~Simon, R.~Penrose, and D.~Bouwmeester, Towards Quantum Superpositions of a Mirror, ibid. 91 (2003) 130401. https://doi.org/10.1103/PhysRevLett.91.130401.

\bibitem{Qin19Proposal}
W.~Qin, A.~Miranowicz, G.~L.~Long, J.~Q.~You, and F.~Nori, Proposal to test quantum wave-particle superposition on massive mechanical resonators, npj Quantum Inf. 5 (2019) 58.
https://doi.org/10.1038/s41534-019-0172-9.

\bibitem{Sun21Remote}
F.-X.~Sun, S.-S.~Zheng, Y.~Xiao, Q.~H.~Gong, Q.~Y.~He, and K.~Xia, Remote Generation of Magnon Schr\"{o}dinger Cat State via Magnon-Photon Entanglement, Phys. Rev. Lett. 127 (2021) 087203. https://doi.org/10.1103/PhysRevLett.127.087203.

\bibitem{Aspelmeyer14}
M.~Aspelmeyer, T.~J.~Kippenberg, and F.~Marquardt, Cavity optomechanics, Rev. Mod. Phys. 86 (2014) 1391--1452. https://doi.org/10.1103/RevModPhys.86.1391.

\bibitem{Murch08Observation}
K.~W.~Murch, K.~L.~Moore, S.~Gupta, and D.~M.~Stamper-Kurn, Observation of quantum-measurement backaction with an ultracold atomic gas, Nature Phys. 4 (2008) 561--564. https://doi.org/10.1038/nphys965.

\bibitem{Brennecke08Cavity}
F.~Brennecke, S.~Ritter, T.~Donner, and T.~Esslinger, Cavity optomechanics with a Bose-Einstein condensate, Science 322 (2008) 235--238. https://doi.org/10.1126/science.1163218.

\bibitem{Smith11Optomechanical}
M.~H.~Schleier-Smith, I.~D.~Leroux, H.~Zhang, M.~A.~Van~Camp, and V.~Vuleti\'{c}, Optomechanical Cavity Cooling of an Atomic Ensemble, Phys. Rev. Lett. 107 (2011) 143005. https://doi.org/10.1103/PhysRevLett.107.143005.

\bibitem{Spethmann16Cavity}
N.~Spethmann, J.~Kohler, S.~Schreppler, L.~Buchmann, and D.~M.~Stamper-Kurn, Cavity-mediated coupling of mechanical oscillators limited by quantum back-action, Nat. Phys. 12 (2016) 27--31. https://doi.org/10.1038/nphys3515.

\bibitem{Purdy10Tunable}
T.~P.~Purdy, D.~W.~C.~Brooks, T.~Botter, N.~Brahms, Z.-Y.~Ma, and D.~M.~Stamper-Kurn, Tunable Cavity Optomechanics with Ultracold Atoms, Phys. Rev. Lett. 105 (2010) 133602. https://doi.org/10.1103/PhysRevLett.105.133602.

\bibitem{Jing11Quantum}
H.~Jing, D.~S.~Goldbaum, L.~Buchmann, and P.~Meystre, Quantum Optomechanics of a Bose-Einstein Antiferromagnet,  Phys. Rev. Lett. 106 (2011) 223601. https://doi.org/10.1103/PhysRevLett.106.223601.

\bibitem{Kumar21Cavity}
P.~Kumar, T.~Biswas, K.~Feliz, R.~Kanamoto, M.-S.~Chang, A.~K.~Jha, and M. Bhattacharya, Cavity Optomechanical Sensing and Manipulation of an Atomic Persistent Current, Phys. Rev. Lett. 127 (2021) 113601. https://doi.org/10.1103/PhysRevLett.127.113601.

\bibitem{Singh12Quantum}
S.~Singh, H.~Jing, E.~M.~Wright, and P.~Meystre, Quantum-state transfer between a Bose-Einstein condensate and an optomechanical mirror, Phys. Rev. A 86 (2012) 021801(R). https://doi.org/10.1103/PhysRevA.86.021801.

\bibitem{Dong11Multistability}
Y.~Dong, J.~W.~Ye, and H.~Pu, Multistability in an optomechanical system with a two-component Bose-Einstein condensate, Phys. Rev. A 83 (2011) 031608(R). https://doi.org/10.1103/PhysRevA.83.031608.

\bibitem{Zhang12Role}
K.~Y.~Zhang, P.~Meystre, and W.~P.~Zhang, Role Reversal in a Bose-Condensed Optomechanical System, Phys. Rev. Lett. 108 (2012) 240405. https://doi.org/10.1103/PhysRevLett.108.240405.

\bibitem{Inouye98Observation}
S.~Inouye, M.~R.~Andrews, J.~Stenger, H.-J.~Miesner, D.~M.~Stamper-Kurn, and W.~Ketterle, Observation of Feshbach resonances in a Bose-Einstein condensate, Nature (London) 392 (1998) 151--154;  https://doi.org/10.1038/32354.
P.~Weckesser, F.~Thielemann, D.~Wiater, A.~Wojciechowska, L.~Karpa, K.~Jachymski, M.~Tomza, T.~Walker, and T.~Schaetz, Observation of Feshbach resonances between a single ion and ultracold atoms, ibid. 600 (2021) 429--433. https://doi.org/10.1038/s41586-021-04112-y.

\bibitem{Shahmoon20Cavity}
E.~Shahmoon, D.~S.~Wild, M.~D.~Lukin, and S.~F.~Yelin, Cavity quantum optomechanics with an atom-array membrane,  arXiv: 2006.01973 (2020).

%\bibitem{Rabl11Photon}
%P.~Rabl, Photon blockade effect in optomechanical systems, Phys. Rev. Lett. 107 (2011) 063601. doi: 10.1103/PhysRevLett.107.063601.

\bibitem{Baumann10Dicke}
K~Baumann, C.~Guerlin, F.~Brennecke, and T.~Esslinger, Dicke quantum phase transition with a superfluid gas in an optical cavity, Nature (London) 464 (2010) 1301--1306. https://doi.org/10.1038/nature09009.

\bibitem{Stenger98Spin}
J.~Stenger, S.~Inouye, D.~M.~Stamper-Kurn, H.-J.~Miesner, A.~P.~Chikkatur, and W.~Ketterle, Spin domains in ground-state Bose-Einstein condensates, Nature (London) 396 (1998) 345--348. https://doi.org/10.1038/24567.

\bibitem{Myatt97Production}
C.~J.~Myatt, E.~A.~Burt, R.~W.~Ghrist, E.~A.~Cornell, and C.~E.~Wieman, Production of Two Overlapping Bose-Einstein Condensates by Sympathetic Cooling, Phys. Rev. Lett. 78 (1997) 586--589. https://doi.org/10.1103/PhysRevLett.78.586.

\bibitem{Kuang07Generation}
L.-M.~Kuang, Z.-B.~Chen, and J.-W. Pan, Generation of entangled coherent states for distant Bose-Einstein condensates via electromagnetically induced transparency, Phys. Rev. A 76 (2007) 052324. https://doi.org/10.1103/PhysRevA.76.052324.

\bibitem{Greiner03Emergence}
M.~Greiner, C.~A.~Regal, and  D.~S.~Jin, Emergence of a molecular Bose-Einstein condensate from a Fermi gas, Nature (London) 426 (2003) 537--540. https://doi.org/10.1038/nature02199.

\bibitem{Morsch06Dynamics}
O.~Morsch and M.~Oberthaler, Dynamics of Bose-Einstein condensates in optical lattices, Rev. Mod. Phys. 78 (2006) 179--215. https://doi.org/10.1103/RevModPhys.78.179.

\bibitem{Verhagen12Quantum}
E.~Verhagen, S.~Del\'{e}glise, S.~Weis, A.~Schliesser, and T.~J.~Kippenberg, Quantum-coherent coupling of a mechanical oscillator to an optical cavity mode, Nature (London) 482 (2012) 63--67. https://doi.org/10.1038/nature10787.
	
\bibitem{Kuang03Generation}
L.-M.~Kuang and L.~Zhou, Generation of atom-photon entangled states in atomic Bose-Einstein condensate via electromagnetically induced transparency, Phys. Rev. A 68 (2003) 043606. https://doi.org/10.1103/PhysRevA.68.043606.

\bibitem{Walls94Quantum}
D.~F.~Walls and G.~J.~Milurn, \textit{Quantum Optics} (Spinger-Verlag, Berlin, 1994).

\bibitem{Savchenkov07Optical}
A.~A.~Savchenkov, A.~B.~Matsko, V.~S.~Ilchenko, and L.~Maleki, Optical resonators with ten million finesse, Opt. Express 15(11) (2007) 6768--6773. https://doi.org/10.1364/OE.15.006768.

\bibitem{Honari21Fabrication}
S.~Honari, S.~Haque, and T.~Lu, Fabrication of ultra-high Q silica microdisk using chemo-mechanical polishing,  Appl. Phys. Lett. 119 (2021) 031107. https://doi.org/10.1063/5.0051674.

\bibitem{Qin18Exponentially}
W.~Qin, A.~Miranowicz, P.-B.~Li, X.-Y.~L\"{u}, J.~Q.~You, and F.~Nori, Exponentially Enhanced Light-Matter Interaction, Cooperativities, and Steady-State Entanglement Using Parametric Amplification, Phys. Rev. Lett. 120 (2018) 093601. https://doi.org/10.1103/PhysRevLett.120.093601.

		
\end{thebibliography}
\end{document}